\documentclass[prb,twocolumn,10pt,aps]{revtex4-2}
\usepackage[utf8]{inputenc}
\usepackage{textcomp}
\usepackage{amssymb}
\usepackage{natbib}
\usepackage{amsmath}
\usepackage{amsfonts}
\usepackage{graphicx}
\usepackage{subfigure}
\usepackage{mathrsfs}
\usepackage{dcolumn} 
\usepackage[T1]{fontenc}
\usepackage{bm}
\usepackage{color}
\usepackage{lipsum}
\usepackage{cancel}
\usepackage{mathbbol}
\usepackage{epsfig}
\usepackage{units}
\usepackage{esint}
\usepackage{soul} 
\usepackage{url}
\usepackage{afterpage}
\usepackage{hyperref}
\usepackage{braket}
\newcommand{\me}[1]{\left\langle #1 \right\rangle }

\newcommand{\tr}{\mathrm{tr}}

\begin{document}
	\title{On the Conditions for a Quantum Violent Relaxation} 
	\author{Guido Giachetti$^{1,2}$}
	\author{Nicol\`o Defenu$^{3,4}$}
	\affiliation{$^{1}$ CY Paris Universit\'e, 2 Av. Adolphe Chauvin, Pontoise, France}
        \affiliation{$^{2}$ Laboratoire de Physique de l'\'Ecole Normale Sup\'erieure, ENS $\&$ PSL University, 75005 Paris, France}
	\affiliation{$^{3}$Institut f\"ur Theoretische Physik, ETH Z\"urich, Wolfgang-Pauli-Str. 27 Z\"urich, Switzerland}
    \affiliation{$^{4}$CNR-INO, Area Science Park, Basovizza, I-34149 Trieste, Italy}
	\begin{abstract}
	\noindent
    While the dynamics of fully-connected systems is dominated my mean-field effect, in the classical limit the single-particle observables are observed to relax toward a non-thermal stationary value. This phenomenon is known as Violent Relaxation and it is general absent at the quantum mechanical level, where single-particle observables exhibits long-lived oscillations. In this paper we explain this discrepancy by determining some very restrictive condition that quantum single-site Hamiltonian should meet in order for the system to undergo relaxation. We thus check them by introducing a new model (the so-called $w$-model) which can exhibit dynamical phase transition between a thermal behavior, persistent excitations and violent relaxation. We also propose a way to implement it within light-matter coupling. We finally explain how the classical limit restore this behavior, thus showing that, even if the mean-field dynamics of quantum models is usually thought to be classical, quantum effects still play an important role in it.    
  \end{abstract}
	\maketitle 
\noindent
\emph{Introduction:} 
The physics of long-range interacting systems, i.e. those systems whose microscopic components are coupled through slow-decaying power-law potentials $\sim r^{-\alpha}$\,\cite{campa2009statistical, campa2014physics}, has recently experienced a new wave of interest due to their possible realization in multiple experimental setups\,\cite{defenu2021longrange}. Depending on the decay exponent $\alpha$, long-range interactions generate several novel features in classical physics, such as tuneable critical scaling in the weak long-range regime ($\alpha>d$)\,\cite{sak1973recursion,Kosterlitz76,defenu2015fixed,Giachetti2021} or ensemble inequivalece\,\cite{campa2009statistical,barre2001inequivalence, barre2005large} and quasi-stationary states\,\cite{sylos2012violent} in the strong long-range regime ($\alpha<d$). Quantum Long-range physics has been shown to be even richer, displaying plenty of peculiar features such as anomalous defect scaling\,\cite{acevedo2014new,defenu2018dynamical,safavi2018verification,keesling2019quantum}, long-lived prethermal behavior\,\cite{kastner2011diverging,defenu2021metastability,schutz2014prethermalization}, dynamical phase transitions\,\cite{zhang2017observation, baumann2010dicke, das2006infinite,sciolla2010quantum,sciolla2013quantum,vzunkovivc2016dynamical}, anomalous entanglement scaling\,\cite{pappalardi2018scrambling,lerose2020origin,giachetti2021entanglement}, Floquet time-crystals\,\cite{rovny2018observation, choi2017observation, zhang2017observation,RussomannoPRB2017,surace2019floquet,giachetti2022high}, and several others\,\cite{defenu2021longrange,defenu2024out} which have proven to be robust against external noise and competing short-range perturbations\,\cite{kastner2011diverging,lerose2018chaotic,defenu2021metastability,piccitto2021symmetries}. 

In particular, fully connected Hamiltonians ($\alpha =0$) can be engineered in cavity experiments\,\cite{schutz2014prethermalization,mivehvar2021cavity} or Rydberg gases\,\cite{lahaye2009physics,chomaz2023dipolar} or seen as a large-dimension approximation of short-range models\,\cite{metzner1980correlated,eckstein2009thermalization}. As the mean-field description becomes in general exact in the thermodynamic limit $N \rightarrow \infty$\,\cite{pearce1975anisotropic,pearce1978high,tindemans1974exact,kastner2010nonequivalence}, the dynamics of quantum fully-connected systems can be usually reduced to a small number of classical macroscopic degrees of freedom in the thermodynamic limit\,\cite{sciolla2010quantum,sciolla2013quantum}, resulting in long-lived periodic oscillations of the order parameter\cite{blass2018quantum,igloi2018quantum} (the picture can also be easily generalized to most strong-long range systems\,\cite{defenu2021metastability}). This is the case, e.g., of the Lipkin-Meshkov-Glick (LMG) model, \textcolor{black}{originally introduced in the context of nuclear physics \cite{lipkin1965validity} and become the paradigmatic example for fully-connected model, as an all-to-all interacting spin model \cite{caneva2008,RibeiroPRE2008,campbell2016,sela2016molecular,vzunkovivc2016dynamical,pappalardi2018scrambling,PizziNatComm2021,Munozarias2022}.} 

Also in the classical case, mean-field effects prevent the loss of memory and the onset of the thermalization of fully-connected systems up to a time-scale $O(N^\zeta)$ (for some $\zeta >0$). In spite of this, classical all-to-all models exhibit a remarkably different behavior with respect to their quantum counterpart: indeed, instead of exhibiting long-lived excitations, generically, large-scale observables are observed to asymptotically relax toward an equilibrium value, which does not correspond with the thermal one, on a finite size-independent time scale. This phenomenon is known as Violent Relaxation (VR)\,\cite{Tremaine1986,campa2009statistical} and it is also believed to be a key ingredient to explain the formation of large-scale astrophysical structures\,\cite{henon1964evolution, lyndenbell1967statistical,vanAlbada1982,chavanis1996statistical,sylos2012violent,chavanis1998degenerate,chavanis2005coarse,chavanis2006coarse}. 
VR has been extensively studied, particularly in the paradigmatic Hamiltonian-Mean-Field (HMF) model\,\cite{antoni1995clustering}. 

Non-thermal relaxation phenomena are known to occur at the crossover between integrability and chaos, both in classical\,\cite{cao2018incomplete} and quantum\,\cite{yurovsky2011memory} realms. However, VR is substantially different, as the system becomes integrable only later due to relaxation itself, while scrambling plays a crucial role in the initial dynamics\,\cite{lyndenbell1967statistical,chavanis1996statistical,giachetti2020coarse,chavanis2006lynden,chavanis2006coarse,giachetti2019violent,santini2022violent}. We note that no general framework comparable to the H-theorem in usual kinetic theory\,\cite{kennard1938kinetic}, or the thermalization to a generalized Gibbs ensemble\,\cite{rigol2007relaxation,Rigol2008} in the integrable case, has yet emerged to quantitatively predict VR.

The discrepancy between the classical relaxation and the quantum persistent oscillation  cannot traced back to the presence of quantum correlations: while it is known that initializing a quantum system in a correlated state leads to the violation of mean-field dynamics\,\cite{queisser2014equilibration, yurovsky2017dissociation}; VR is a purely mean field phenomenon, which takes place, in the classical limit, even for uncorrelated initial conditions. This suggests that quantum effects may have a dramatic influence already for factorized ``classical-like" initial state. This is even more surprising due to the fact that mean-field dynamics is expected to be fundamentally classical in its nature. 

Quantum versions of the HMF model have been introduced both in the fermionic\,\cite{chavanis2011quantumI} and bosonic\,\cite{chavanis2011quantumII} cases, and the possibility of violent-relaxation in the latter have been recently investigated in Ref.\,\cite{plestid2018violent}. In spite of these preliminary results, a general picture to explain the onset of violent-relaxation in the quantum world is still missing. This paper is aimed at closing this gap, by addressing the matter in its generality, and by finding the conditions under which fully-connected models undergo VR. In particular we find that such a possibility is closely linked to the property of the spectrum of the mean-field effective Hamiltonian. This result forms a bridge between the classical and the quantum realms, highlighting how single-particle quantum effects affect out-of-equilibrium phenomena. 

The paper is structured as follows: After introducing the general setting in Sec.\,\ref{sec:General}, we derive the main result of the paper, namely the condition for quantum VR in Sec.\,\ref{sec:Condition}. To validate our result, we first examine the fully-connected rotor model in Sec.\,\ref{sec:HMF}, which is analogous to the quantum HMF model. Subsequently, we introduce a new model, the so-called $w$-model, in Sec.\,\ref{sec:wmodel}, which can exhibit either quantum-like or classical-like behavior depending on the choice of parameters. Finally, in Sec.\,\ref{sec:classical}, we discuss how violent relaxation can be recovered for a generic system in the $\hbar \rightarrow 0$ limit.

\section{General setting} \label{sec:General}
Let us consider a generic set of $N$ of quantum systems coupled by a fully-connected, two-body, interaction which can be described by the Hamiltonian 
\begin{equation} \label{eq:H}
    H = \sum_j H_0 (\boldsymbol{\tau}_j) - \frac{\lambda}{2N} \sum_{j,j^\prime} H^a_1 (\boldsymbol{\tau}_j) H^a_1 (\boldsymbol{\tau}_{j^\prime}) 
\end{equation}
where $\bold{\tau}_j$ denotes the set of quantum variables relative to the site $j$, such that $[\boldsymbol{\tau}_j,\boldsymbol{\tau}_l] = 0$ for $i \neq l$, the summation over the index $a$ is implied, and the $1/N$ factor is chosen in agreement with the so-called Kac scaling, which ensures an extensive energy\,\cite{kac1963van}. The interaction is chosen to be ferromagnetic ($\lambda>0$). The $\boldsymbol{\tau}$ variables can be chosen to be spins, canonical conjugate pairs, or rotor variables, but this is not essential for our results. 

Many of the fully-connected models studied in the past can be seen as particular instances of the Hamiltonian in Eq.\,\eqref{eq:H}: for example the LMG model\,\cite{caneva2008} is recovered with the choice $H_0(\mathbf{s}) = - h s_z$ and $H^x_1 = s_x$, $H^y_1 = \sqrt{\gamma} s_y$. 

As previously mentioned, our goal is to derive a comprehensive understanding of how VR emerges from quantum systems in the classical limit. To achieve this, we consider factorized "classical" initial states of the form
\begin{equation}
\label{init_cond}
    \varrho(0) = \frac{1}{N} \sum^N_{j} \rho_j (0)   \, .
\end{equation}
Since quantum correlations are not relevant in the classical limit, the above assumption won't hinder our conclusions. Additionally, it's worth noting that experimental studies of quantum dynamics often focus on uncorrelated initial states due to the relative ease of their preparation\,\cite{monroe2021programmable, defenu2021longrange, chomaz2023dipolar}.  

Under the assumption in Eq.\,\eqref{init_cond}, the mean-field description of the system becomes exact\,\cite{pearce1975anisotropic,pearce1978high,tindemans1974exact,kastner2010nonequivalence,sciolla2010quantum, sciolla2013quantum}. This means that that the dynamics of the expectation values of single-site operators can be described in terms of a single site effective density operator $\varrho(t)$, see Appendix\,\ref{app:meanfield}, which evolves according to the equation of motion
\begin{equation} \label{eq:vonneuman}
    i \frac{d \varrho}{dt} = [\mathcal{H}(t), \rho] \equiv \mathcal{L}_{\mathcal{H}} (\varrho)
\end{equation}
where $\mathcal{H}(t)$ effectively describes the dynamical evolution of a single site
\begin{equation} \label{eq:mathcalH}
    \mathcal{H} (t) = H_0 (\mathbf{\tau}) - \lambda \mu^a (t) H^a_1(\mathbf{\tau}) , 
\end{equation}
and the time-dependent coupling coincides with the order parameter $\mu^a(t)$. This is determined by the self-consistent condition
\begin{equation} \label{eq:selfconst}
    \mu^a(t) = \tr{\left( \varrho (t) H^a_1  \right) } \ .
\end{equation}
For finite $N$, the above picture breaks down on timescales polynomial in $N$, see Ref.\,\cite{pappalardi2018scrambling}. If $\mu(t) = 0$, however, the many-body fluctuations dominates over the expectation value of $H_1$ regardless of the value of $N$, so that the mean-field approximation is no longer valid. Thus, in the symmetry-broken phase, the size of the system $N$ acts as a control parameter for the stability of the semi-classical initial states in Eq.\,\eqref{init_cond}.

Let us notice that the dynamics conserves the quantity $\epsilon = \tr(\varrho(t) H_0) - \lambda \boldsymbol{\mu}^2(t)/2$, see Appendix\,\ref{app:meanfield}, which has the physical dimensions of an energy density. 

\section{Conditions for a relaxation}
\label{sec:Condition}
As the dynamics described by Eqs.\,\eqref{eq:mathcalH} and\,\eqref{eq:selfconst} is effectively one-body, the long-time behavior of the observable is not expected to converge to the thermal average. In analogy with the classical case, VR shall occur if $\lim_{t\to\infty} \boldsymbol{\mu} (t) \rightarrow \mu_{\infty}$ which will differ from the thermal expectation value $\boldsymbol{\mu}_{\infty}\neq \boldsymbol{\mu}_{\rm th}$. While this generically expected in classical long-range physics, in quantum long-range systems this is rather exceptional as long-lived oscillations are expected to survive at long time (this is the case, for example, of the magnetization in the LMG model).

We are going to show that the difference between these two behaviors boils down to the fact that the levels of the single particle Hamiltonian\,\eqref{eq:mathcalH} can exhibit a finite gap, and this hinders the possibility of a relaxation. While a detailed proof of the above statement is provided below, the physical interpretation of this result is straightforward: if one supposes that VR really takes place, $\mathcal{H}(t)$ will become time-independent at large times. Then, the existence of a finite gap would result in infinitely-lived Rabi oscillations between the populations of the energy levels, which contradict our hypotheses. As a consequence, any system with a finite Hilbert space per site (e.g., in the case of spin variables, for any finite $s$) will not undergo VR. While this behavior is indeed expected, it is surprising to notice how, even at the level of mean-field dynamics, for single-site observables, quantum and classical dynamics can exhibit such a dramatically different behavior already at the qualitative level. 
\begin{figure}
    \centering
    \includegraphics[width=
    \columnwidth]{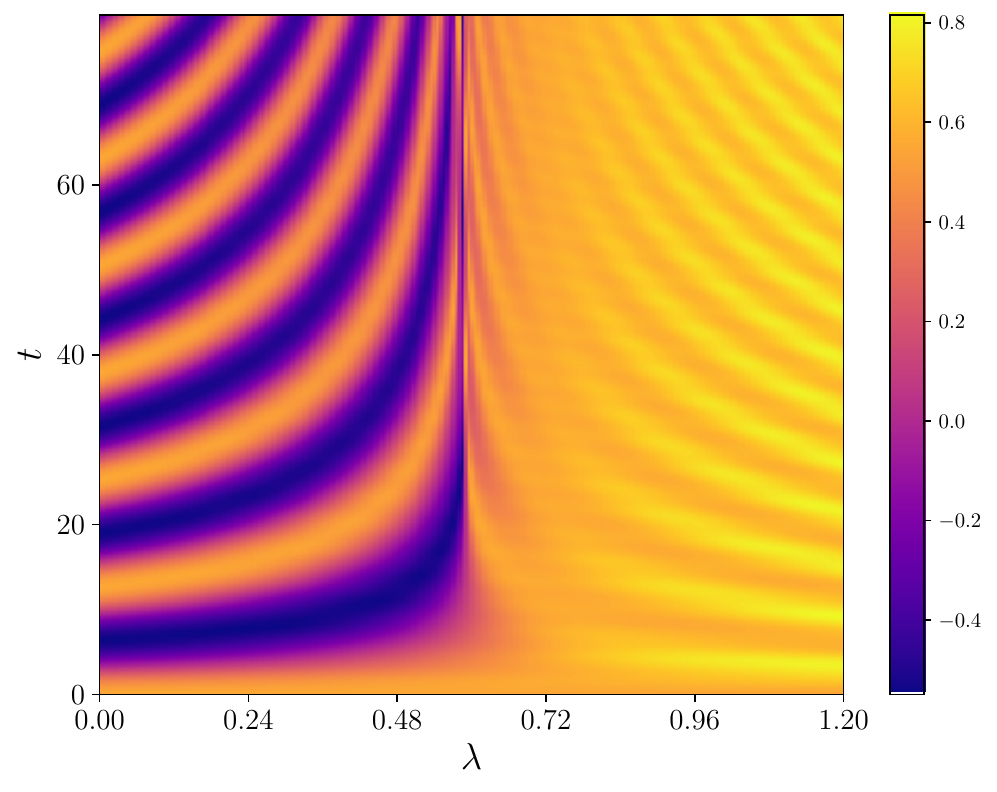}
    \caption{Color plot of the of the magnetization $\mu(t)$ of the two-component quantum rotor model, as a function of the coupling $\lambda$ for a given pure initial state  $\rho(0) = \ket{\Psi(0)} \bra{\Psi(0)}$, with $\ket{\Psi(0)} \propto 3 \ket{L=0} + \ket{L=-1} + \ket{L=1} $. Although the quantitative features of the dynamics of $\mu(t)$ exhibit a sharp transition around $\lambda = 0.6$, it is clear from the figure that no VR actually occur for any value for $\lambda$.}
    \label{fig:QHMf}
\end{figure}
Let us now state rigorously our result, which is the main result of this work, and  give a formal proof.  
\\\\
\textit{A necessary condition for the system described by mean-field dynamics to undergo violent relaxation is that the (dynamically accessible) point spectrum of the single-site Hamiltonian $\mathcal{H}(t)$ in Eq.\,\eqref{eq:mathcalH} contains, at most, a single element (possibly degenerate). In case some of the corresponding states are not accessible to the dynamics because of symmetry reason, those should not be included in the above analysis. If the discrete spectrum is empty, then $\mu_{\infty} = 0$.}
\, \\\\
\emph{Proof:} Let us thus suppose that VR indeed occurs, i.e. that in the limit $t \rightarrow \infty$, $\boldsymbol{\mu}(t)\rightarrow \boldsymbol{\mu}_{\infty}$. As a consequence, at large times the effective Hamiltonian $\mathcal{H}$ in Eq.\,\eqref{eq:mathcalH} can be considered a time-independent operator $\mathcal{H}_{\infty}$, and the evolution \eqref{eq:vonneuman} becomes
\begin{equation}
    i \frac{d}{dt} \varrho(t) = \left[ \mathcal{H}_{\infty}, \varrho (t) \right] \equiv \mathcal{L}_{\infty} (\varrho (t)) .
\end{equation}
Let us now suppose the spectrum $\lbrace \mathcal{E} \rbrace$ of $\mathcal{H}_{\infty}$ is continuous except for a finite set of discrete points $\lbrace \mathcal{E}_k \rbrace$, $k=1, \cdots, n_{\rm b}$. The spectrum of the (time-independent) von Neumann superoperator $\mathcal{L}_{\infty}$ is given by 
\begin{equation} \label{eq:qspectrum}
    \lbrace  \Omega \rbrace = \lbrace \mathcal{E} - \mathcal{E}^{\prime} \rbrace \, ,
\end{equation}
where $\mathcal{E}$, $\mathcal{E}^{\prime}$ belong to the spectrum of $\mathcal{H}_{\infty}$. In particular, the $\Omega$ will be continuous as well, but for a finite set of values of the form $\mathcal{E}_k - \mathcal{E}_{k'}$, with $k,k' = 1, \cdots n_{\rm b}$. We can thus write that, at large times
\begin{equation} \label{eq:rhot}
    \varrho (t) = \int d \Omega  \, \varrho (\Omega) \, d(\Omega) e^{i \Omega t} + \sum^{n_{\rm b}}_{k,k'=1} \varrho_{k,k'} e^{i(\mathcal{E}_k - \mathcal{E}_{k'}) t}
\end{equation}
where $d(\Omega)$ is the density of states of the continuous part of the spectrum, $\rho(\Omega)$, $\rho_{k,k'}$ are operator fixed by the initial conditions and the small-time dynamics. Finally, 
\begin{equation} \label{eq:mut}
\begin{split}
    \mu^a(t) =  &\int d \Omega  \, \text{tr} (\varrho(\Omega) H^a_1) \, d(\Omega) e^{i \Omega t} \\ &+ \sum^{n_{\rm b}}_{k,k'=1} \text{tr} (\varrho_{k,k'} H^a_1) \, e^{i(\mathcal{E}_k - \mathcal{E}_{k'}) t} \ .
\end{split}
\end{equation}
\noindent
By the Riemann-Lebesgue lemma, the first term in the right-hand-side of Eq.\,\eqref{eq:mut} goes to zero in the $t \rightarrow \infty$ limit. This means that, for $n_{\rm b}=0$, $\boldsymbol{\mu}(t) \searrow 0$ at later times. Let us consider now the $n_{\rm b} > 1$ case: the sum in the right-hand-side will contain oscillating terms thus contradicting the hypothesis of a constant $\mu_{\infty}$ in the large time limit. We conclude that we can have VR only for $n_{\rm b}=0$ or $n_{\rm b}=1$, while, for $n_{\rm b}>1$, $\mu(t)$ continues to oscillate. In particular, for $n_{\rm b}=0$, we have that $\mu(t) \rightarrow 0$, while for $n_{\rm b}=1$ one has $\mu_{\infty}\geq 0$. 

Notice that, for $n_{\rm b}=2$, only one harmonic is present on the right-hand side of Eq.\,\eqref{eq:mut}. Nevertheless, this does not imply that $\mu(t)$ is periodic. Indeed, if $\mu(t)$ were periodic at late times, the time-independent effective Hamiltonian $\mathcal{H}_{\infty}$ cannot be defined and our entire constructions does not apply. For the same reason, our arguments do not yield a sufficient condition for VR. In fact, even if $n_{\rm b}=1$ or $n_{\rm b}=0$, parametric resonances may occur at any time, hindering relaxation.  
 
\section{The quantum rotor model}
\label{sec:HMF}
In this section we are going to examine a first example of our statement, i.e. the case of quantum rotor model: $N$-component rotor model can be seen as the result of the quantization of a $N$-dimensional rigid rotor, i.e. a particle moving on the surface of a $N$-sphere. They are known to show the same equilibrium universality class of quantum antiferromagnets\,\cite{sachdev1999quantum}. In particular, we will focus on the case $N=2$ (for which the model can be thought as an array of Josephson junctions\,\cite{vojta2006quantum}): in the fully-connected case, the Hamiltonian is given by  
\begin{equation} \label{eq:HQR}
    H_{\rm QR} = \frac{1}{2} \sum_j L^2_j - \frac{\lambda}{2N} \sum_{j,k} \mathbf{n}_j \cdot \mathbf{n}_k
\end{equation}
where $n^{\alpha}_j$ ($\alpha = 1,2$) are the components of a unity vector ($\mathbf{n}_j^2=1$) and $L_j$ the corresponding angular momentum: $[L_k,n^{\alpha}_j] = i \delta_{jk } \epsilon^{\alpha \beta} n^{\beta}_k $, $\epsilon^{\alpha \beta}$ being the completely asymmetric tensor of rank two (we set $\hbar = 1$). This model falls straightforwardly within the class of Eq.\,\eqref{eq:H} ($H_0 = L$, $H_1^a = n^a$). By introducing the angular parametrization $\mathbf{n}_j = (\cos \theta_j, \sin \theta_j)$, with  $[\theta_j,L_k] = i \delta_{jk}$, we can write the single-site Hamiltonian \eqref{eq:mathcalH} as
\begin{equation} 
    \mathcal{H}_{\rm QR} (t) = \frac{L^2}{2} - \lambda \boldsymbol{\mu} (t) \cdot \left(\cos \theta, \sin \theta \right)
\end{equation}
with $\boldsymbol{\mu} (t) = \tr(\varrho(t) \mathbf{n})$ or, in terms of single components, $\mu_1 (t) = \tr(\varrho(t) \cos \theta)$ and $\mu_2 (t) = \tr(\varrho(t) \sin \theta)$. In the following, we will assume a $\theta \rightarrow - \theta$ symmetry in the initial state, so that 
\begin{equation} \label{eq:mathcalHQR}
    \mathcal{H}_{\rm QR} (t) = \frac{L^2}{2} - \mu (t) \cos \theta
\end{equation}
with $\mu(t) = \tr(\varrho(t) \cos \theta)$. The main reason behind the choice of the Hamiltonian \eqref{eq:HQR} is now clear, as in the $\hbar \rightarrow 0$ limit Eq.\,\eqref{eq:mathcalHQR} reduces to the HMF model\,\cite{antoni1995clustering}. Indeed, while the Hamiltonian Eq.\,\eqref{eq:mathcalHQR} does not exactly coincide with the one of the quantum HMF model introduced in Ref.\,\cite{chavanis2011quantumII}, we notice that, for uniform initial conditions, $\varrho(t)$ is a pure state and the evolution of the system can be expressed though the Gross-Pitajevskij formalism as in the bosonic HMF model \cite{chavanis1996statistical,plestid2018violent}.

In spite of the fact that the single-space Hilbert space is infinite dimensional, let us notice that the spectrum of an Hamiltonian of the form \eqref{eq:mathcalHQR} is completely discrete for any value of the $\mu$, so that we expect the quantum system not to undergo VR. Indeed, as shown in Fig.\,\ref{fig:QHMf}, the evolution of $\mu_1 (t)$, obtained by a numerical computation of Eq.\,\eqref{eq:mut}, does not converge to an asymptotic constant value.

\section{The \texorpdfstring{$w$}{w}-model}
\label{sec:wmodel}
In order to see an instance of VR in the quantum world we introduce the $w$-model Hamiltonian, i.e. a fully connected spin model defined by the Hamiltonian
\begin{equation} \label{eq:Hw}
    H_{w} = -\frac{h}{s} \sum_j s^x_j - \frac{\lambda}{2N} \sum_{j,k} \theta_H (w^2 - s^{z2}_j) \theta_H (w^2 - s^{z2}_k)
\end{equation}
which  is $\mathbb{Z}_2$ symmetric ($s_j^z \rightarrow -s_j^z$). In this expression we denote with $\theta_H (x)$ the Heaviside step function; by convention $\theta_H (0) = 1$ and $w< s$ is a non-negative integer. Physically $H_w$ is characterized by the competition between the magnetic field $h$, which act as a hopping term in the space of that magnetic quantum number $m$, and the collective interaction, which tends to localize the system in $|m| \leq w$. 
\begin{figure}
    \centering
    \includegraphics[width= 0.45
    \textwidth]{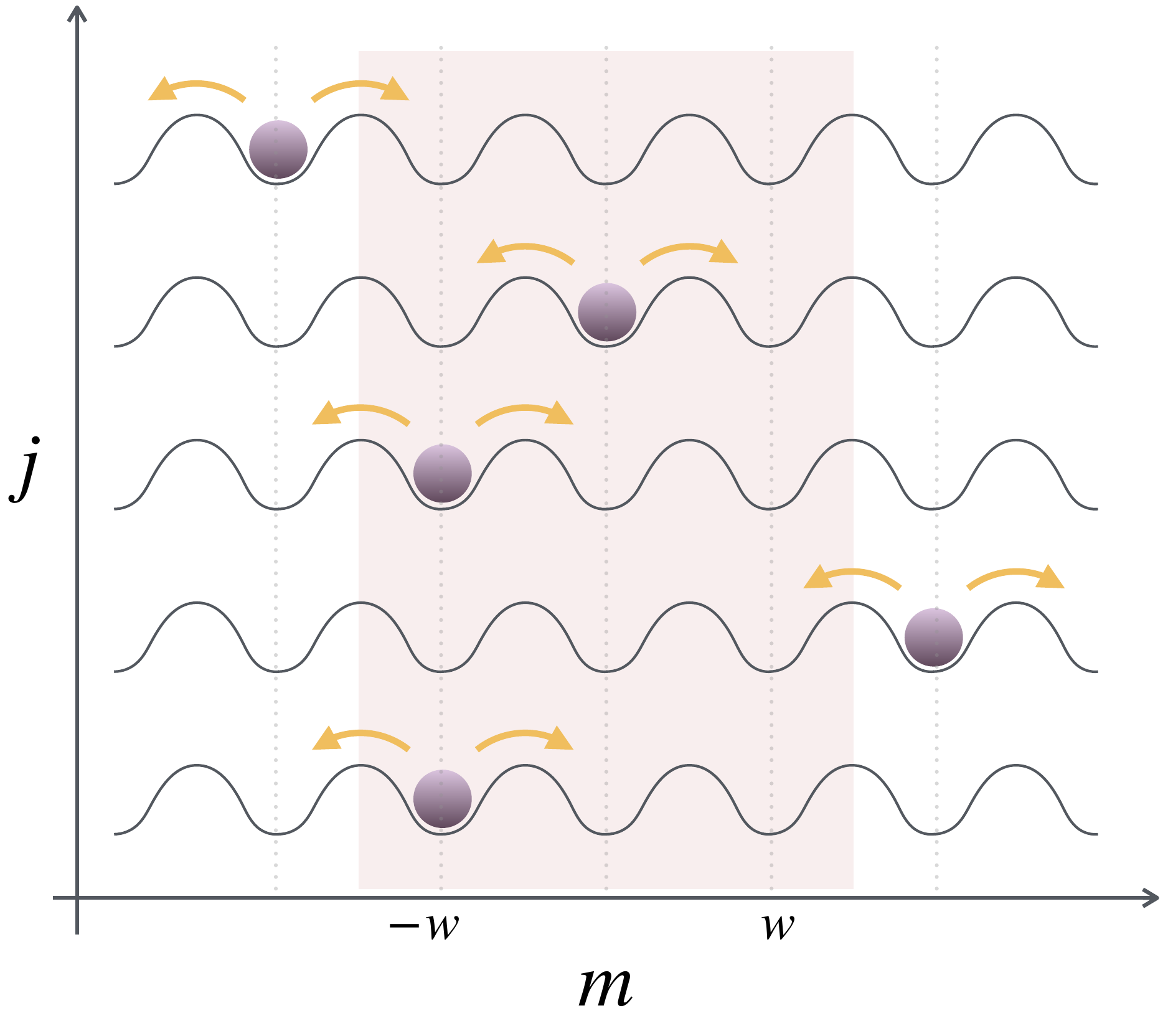}
    \caption{\textcolor{black}{Sketch of the particle interpretation of the $w$-model ($w=1$ in the picture). The particle hops in along the $m$ axis, while within the shaded region (the cavity), corresponding to $|m| \leq w$, they undergo a density-density all-to-all interaction. }}
    \label{fig:OL}
\end{figure}

\subsection{Bosonic interpretation of the \texorpdfstring{$w$}{w}-model}
\noindent
Before examining the dynamics of the system, let us notice that $H_w$ can be alternatively thought as a two-dimensional bosonic lattice Hamiltonian with collective density-density Hubbard-like interactions. In this perspectve, the Hamiltonian describes a square lattice of $(2s+1) \times N$ sites, whose generic site is denoted by $(m,j)$, $m=-s,\dots,s$, $j=1,\dots,N$. We will denote with $a^\dagger_{m,j}$, $a_{m,j}$ the bosonic creation and annihilation operators relative to each site $(m,j)$. We consider a situation in which we have a single particle for any row, namely 
\begin{equation}
    n_j \equiv \sum^{m}_{m=-s} n_{m,j} = 1  \hspace{1cm} \forall \ j \in {1,N} \, ,  
\end{equation}
where $n_{m,j} = a^{\dagger}_{m,j} a_{m,j}$ is the number operator. The particles experience a non-homogeneous, nearest-neighbors hopping in the $m$ direction, while particles in different rows interact through a density-density infinite-range coupling, only in the region $-w \leq m \leq m$, see the sketch in Fig.\,\ref{fig:OL}. The Hamiltonian of the system can be written as
\begin{equation} \label{eq:Hlattice}
\begin{split}
    H^{\rm lattice}_{w} &= - h \sum^{s-1}_{m=-s} (t(m) \ a^\dagger_{m+1,j} a_{m,j} + \text{h.c.}) \\ &- \frac{\lambda}{2N} \sum^N_{j,j^{\prime}=1} \sum^{w}_{m=-w} n_{m,j} n_{m.j^{\prime}} \, .
    \end{split}
\end{equation}
As the Hamiltonian\,\eqref{eq:Hlattice} commutes with all the $n_j$, we can restrict ourselves to the sector $n_j \equiv 1$, which gives a $(2s+1)$-dimensional local Hilbert space for any row $j$. By choosing the appropriate hopping parameter as
\begin{equation}
    t(m) = \frac{1}{2s} \sqrt{s(s+1)-m(m+1)} \, ,
\end{equation}
we have that $H^{\rm lattice}_{w}$ restricted to the $n_j \equiv 1$ subspace reproduces exactly $H_{w}$.

The formulation in Eq.\,\eqref{eq:Hlattice} paves the way for engineering the Hamiltonian in Eq.\,\eqref{eq:Hw} in the context of atomic molecular and optical systems. There, flat interactions can be achieved using resonant cavity modes\,\cite{landig2016quantum,mivehvar2021cavity}, although it could be challenging to spatially confine the collective interaction between $m=-w,\cdots, w$. Alternatively, Eq.\,\eqref{eq:Hlattice} could be realized assuming that the index $j$ represents a synthetic dimension, running over the internal states of the atoms\,\cite{hazzard2023synthetic}, while the collective interaction is realized through coherent light beams. However, the latter setting may limit the reachable size $N$ and make it difficult to reach the mean field limit.

On the contrary, the lattice setting depicted in Fig.\,\ref{fig:OL} allows rather large values of $N$ and $s$. The precise form of the hopping function $t(m)$ is not crucial to investigate VR, which, in our picture, is a universal phenomenon and shall not depend on $t(m)$ for $|m| \gg w$. As a consequence, since in the large-$s$ limit $t(m) \sim 1/2$ for $|m| \leq w$, $t(m)$ could be replace by a constant function, which is easier to realize. 
\begin{figure*}
    \centering
    \includegraphics[width= 0.43
    \textwidth]{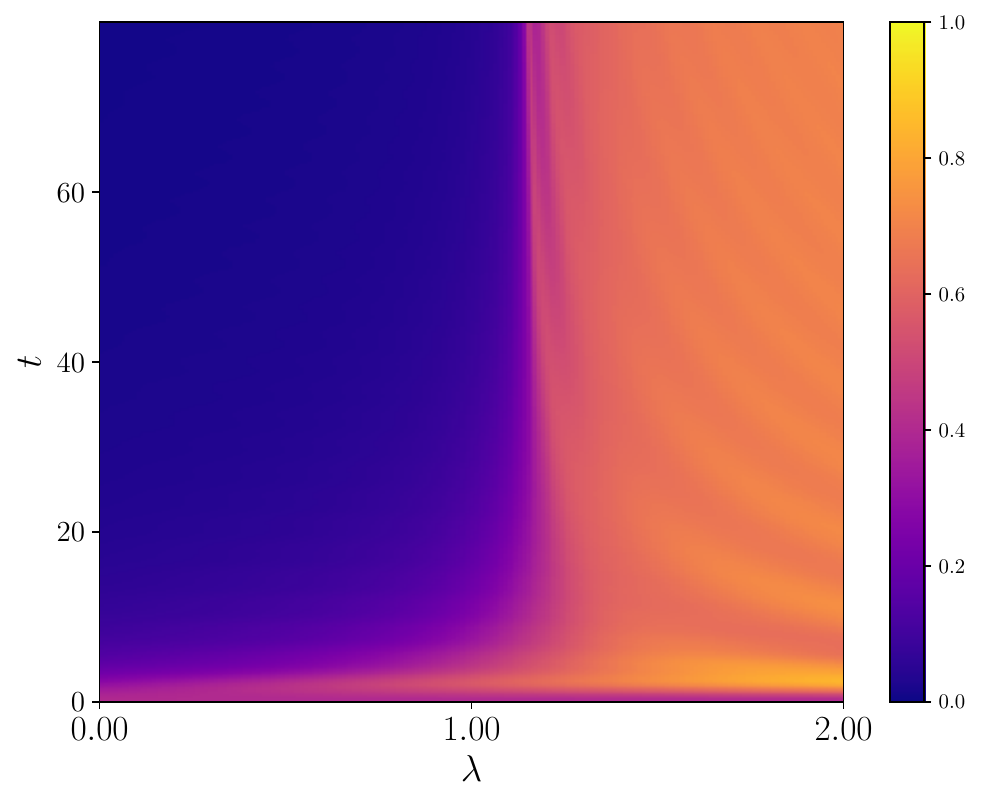}
    \includegraphics[width= 0.43
    \textwidth]{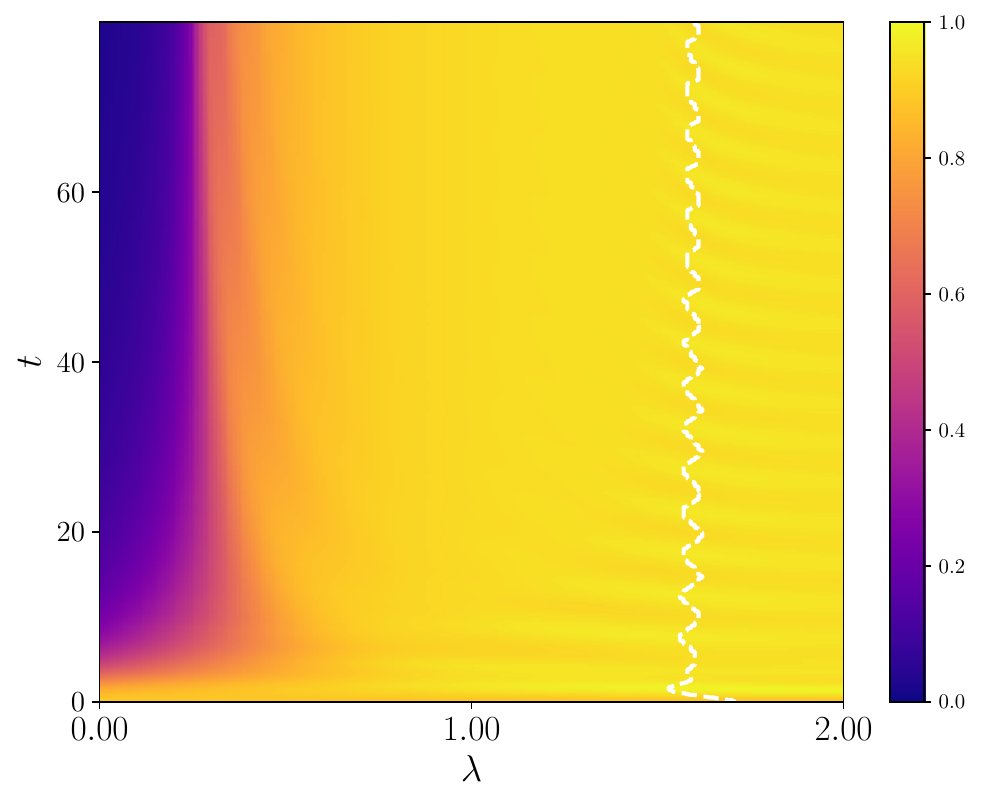} \\
    \includegraphics[width= 0.44
    \textwidth]{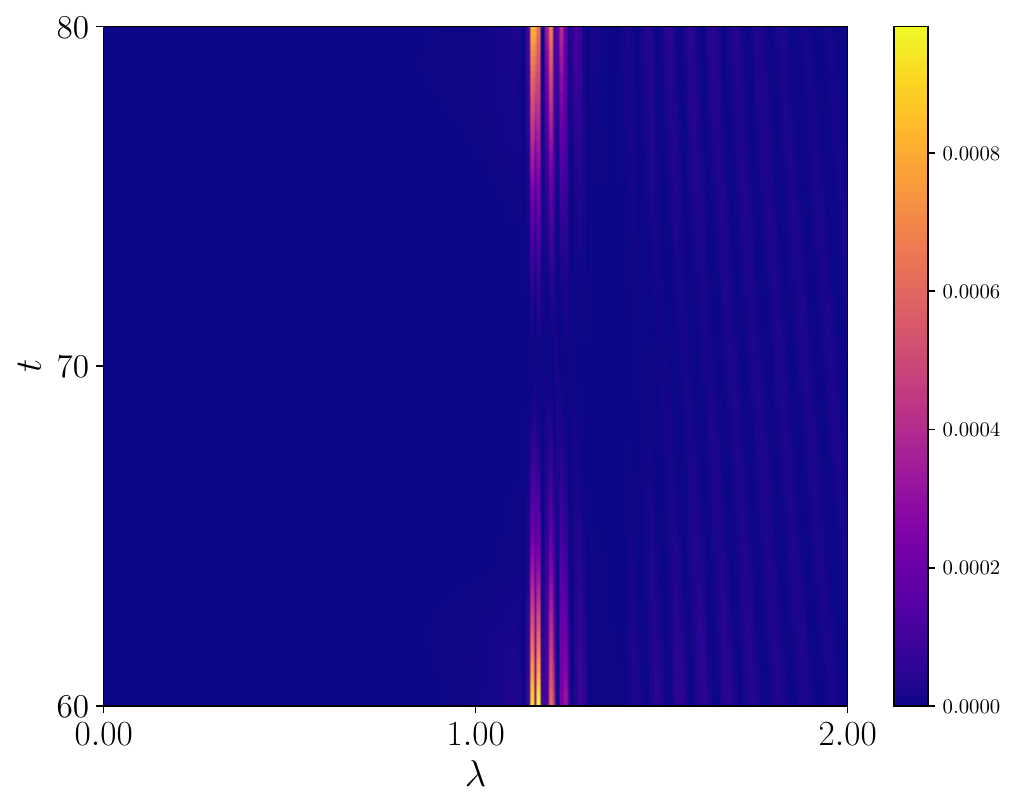}
    \includegraphics[width= 0.44
    \textwidth]{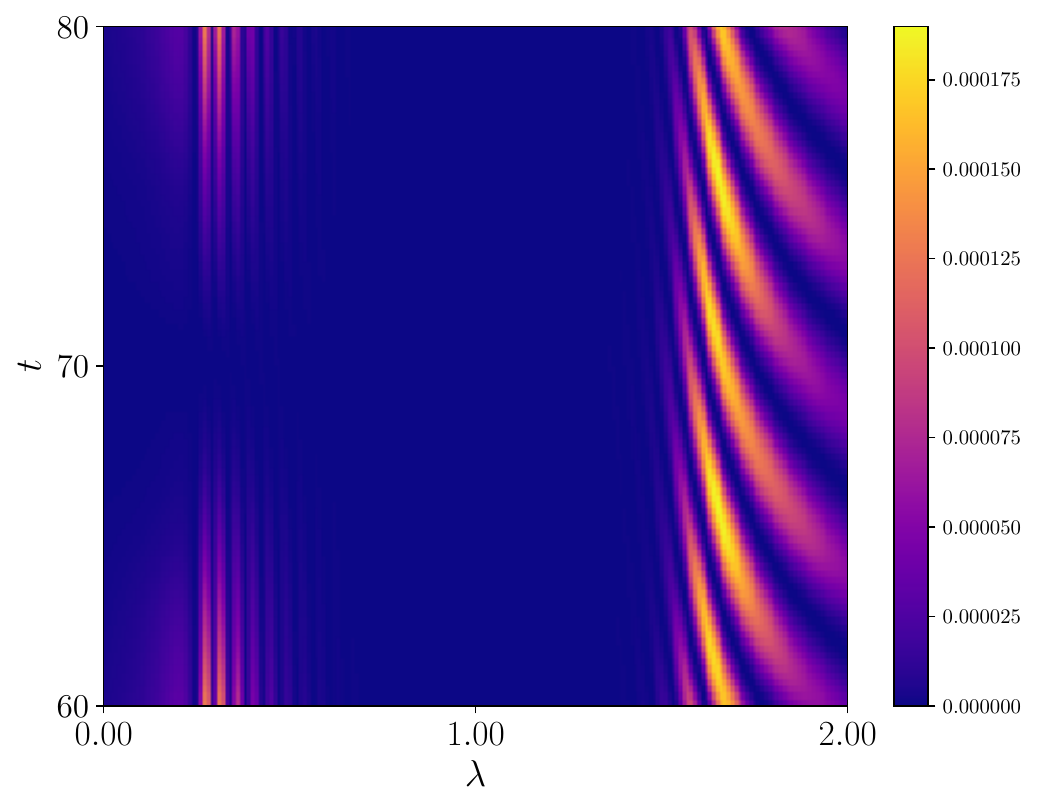}
    \caption{Color plot of the order parameter $\mu(t)$ of the $w$-model (top) and of the oscillation amplitude $A(t)$, defined in Eq.\,\eqref{ampl}, (bottom) as a function of the coupling $\lambda$ for $w=0$ (left) and $w=1$ (right). The parameter chosen are $h=1$, $s=150$, while the initial state is a pure state $\varrho(0) = \ket{\Psi(0)} \bra{\Psi(0)}$ with $\braket{m|\Psi(0)} \propto e^{-m^2/4}$.  As the gap between two contiguous states in the continuous part of the spectrum closes as $s^{-1}$, we expect to see recurrence times as $t \sim 1/s$, so that we investigate $t \lesssim 100$. While for $w=0$ $\mu(t)$ converges to a constant asymptotic value regardless of the value of $\lambda$, for $w=1$ at large enough $\lambda$ the system exhibits persistent oscillation, signaling the presence of two bound-state and the absence of VR. In agreement with our results, the violent-relaxation phase falls within the region $\lambda \mu < x_c(1) \sim 1.51$ (to the left to the white dotted line in the top right panel), in which the single particle Hamiltonian \eqref{eq:mathcalHw} has only one accessible bound states. In both panels, at the left of the VR phase a delocalized thermal phase occurs where $\lim_{t\to\infty}\mu(t)\to 0$. See the text for an extended discussion. }
    \label{fig:Wmodel}
\end{figure*}

\subsection{Dynamics of the model}
Let us now examine the dynamics of the the $w$-model in Eq.\,\eqref{eq:Hw} in the thermodynamic limit ($N \rightarrow \infty$). For each finite $s$, the single-site Hilbert state is finite, since the $s$ spin operators are the higher representation of the $SU(2)$ symmetry group\,\cite{sakurai1999advanced}. However, the single site spectrum become continuous in the $s \rightarrow \infty$ limit possibly resuming VR. 

The single-site Hamiltonian \eqref{eq:mathcalH} corresponding to the $w$-model reads
\begin{equation} \label{eq:mathcalHw}
    \mathcal{H}_w (t) = \frac{h}{s} s_x - \lambda \mu(t) \theta_H (w^2 - s_z^2) , 
\end{equation}
where 
\begin{equation} \label{eq:muw}
\mu(t) = \tr{\left[ \varrho(t) \theta_H (w^2 - s_z^2) \right]}    \, 
\end{equation} 
and $h=1$ has been set for convenience.

We first consider the case $\mu(t) = \bar{\mu} > 0$. As shown in Appendix\,\ref{app:wspectrum}, for any constant $\bar{\mu}$ Hamiltonian\,\eqref{eq:mathcalHw} has a continuum of eigenstates with energy $\mathcal{E} > - 1$, delocalized with respect to the magnetic number $m$, and a finite number of non-degenerate bound states, localized around $m =0$, with a discrete set of energies $\mathcal{E}_n < - 1$. The number $n_{\rm b}$ of the bound states varies from $1$, for $\lambda \bar{\mu} \rightarrow 0^{+}$, till a maximum of $2w+1$ in the limit $\lambda \bar{\mu} \rightarrow + \infty$, their parity with respect to $\mathbb{Z}_2$ ($m \rightarrow - m$) being $(-1)^n$, $n=0,\cdots, n_{\rm b}$. This picture becomes particularly intuitive if we interpret Eq.\,\eqref{eq:Hw} as a bosonic Hamiltonian. The particle hops on a 1-dimensional lattice with a finite square-well potential, which generates the bound states, whose finite number is a direct consequence of the lattice spacing. 

Based on the above analysis, the dynamics induced by Hamiltonian\,\eqref{eq:mathcalHw}, with $\mu(t)$ determined self-consistently by Eq.\,\eqref{eq:muw}, shall display three different qualitative regimes. In the following, we will restrict to $\mathbb{Z}_2$-even initial conditions, so that the maximum number of accessible bound states is $w+1$. For $w>0$ we will denote as $x_{c}=\lambda_{c}\bar{\mu}$ the critical value at which the second even eigenstate appears in the static analysis. Within these conditions the three regimes may be described as follows:
\begin{enumerate}
    \item For small $\lambda$, the potential well is not deep enough to trap the particle and the wave-function eventually delocalizes at large times. Therefore, thermal behavior is recovered and $\lim_{t\to\infty}\mu(t) \rightarrow 0$. The system evolves according to the Hamiltonian $\mathcal{H}_w = s_x/s$ at large times and the spectrum becomes fully continuous. 
    \item For larger $\lambda$, the first (localized) bound state survives at large times. If the oscillations of $\mu(t)$ do not trigger a parametric resonance in the system, $\mu(t)$ relaxes to a finite value. $\lim_{t\to\infty}\mu(t) \rightarrow 0$, consistently with the picture in Sec.\,\ref{sec:Condition}. 
    \item Above a certain critical value $\lambda>\lambda_{c}(w)$ and for $w>1$, the Hamiltonian \eqref{eq:mathcalHw} develops multiple bound-states which survive for any $t>0$. Then, the finite energy gaps will give rise to oscillations which prevent relaxation of $\mu(t)$ toward a constant value. 
\end{enumerate}
The $w$-model represents the prototypical example of the link between spectral properties and quantum VR.

The numerical results shown in Fig.\,\ref{fig:Wmodel} confirm the existence of regimes (1), (2) and (3): for $w=0$ the presence of a single bound-state leads to a dynamical phase transition between a thermal phase ($\mu(t) \rightarrow 0$) at small values of $\lambda$ and a proper VR phase at $\lambda \gtrsim 1.2$. In fact, at small $\lambda$,  $\mu(t)$ vanishes in the long-time limit. In contrast, for $\lambda\gtrsim 1.2$ the system undergoes VR and $\mu(t)$ reaches a stationary non-thermal value $\mu_{\infty}$ at large times. See the left panels in Fig.\,\ref{fig:Wmodel}. Our analysis does not rule out the presence of small, long-lived oscillations in the localized phase which, anyway, would be compatible with the results of Sec.\,\ref{sec:Condition} as only a necessary condition for VR have been established.

For $w>0$ a second even bound state emerges for $\lambda > \lambda_c (w)$. Its emergence, as shown in Fig.\,\eqref{fig:Wmodel} (right panel), results into a new phase for large $\lambda$ in which $\mu(t)$ no longer relaxes; rather it exhibits undamped oscillations. In agreement with the analytical study at constant $\bar{\mu}$, the phase in which the system exhibits violent-relaxation falls within the region of parameter space in which $\mathcal{H}_{w}$ has only one accessible bound states, namely $\lambda \mu < x_c(1) \sim 1.51$ (see Appendix\,\ref{app:wspectrum}).

One can also provide a more qualitative measure of the relaxation, by introducing the amplitude $A(t)$ of the oscillation of $\mu(t)$, defined as 
\begin{equation}
\label{ampl}
    A(t) =  (\mu(t) - \mu_{\infty})^2 \ , 
\end{equation}
where 
\begin{equation}
    \mu_{\infty} = \frac{1}{t_{\rm max} - t_1} \int_{t_1}^{t_{\rm max}} \mu(t) \ dt
\end{equation}
and $t_{\rm max} > t_1 \gg 1$ (for the numerics $t_1 = 60$, $t_{\rm max} = 80$). The values of $A(t)$ as a function of $\lambda$ for $w=0$, $w=1$ are shown in Fig.\,\ref{fig:Wmodel}. As already noticed, while for $w=1$ a region with persistent oscillations develops for large $\lambda$, the phase boundary (both for $w=0$ and $w=1$) is somewhat less clear as close to the phase border some weak, long-wavelength, oscillation seem to persist for large times. While this behavior does not show an appreciable dependence on $s$, ruling out the hypothesis of a finite-size effect, it is not clear whether the oscillations result from a long damping timescale - as the gap between the bound state and the continuum closes - or if they survive in the $t \rightarrow \infty$ limit - as effect of a parametric resonance. Further investigations will be needed to clarify the actual fate of these oscillations.

\section{Classical limit}
\label{sec:classical}
Although VR is the exception rather than the rule, in quantum systems, it is an ubiquitous phenomenon in classical all-to-all interacting models. Such behavior should thus emerge naturally in the classical limit, regardless of the single-particle spectrum of the quantum problem.  

In order to demonstrate the above statement, let us assume the system under study to have well-defined classical limit. We resort to the Wigner function formalism\,\cite{wigner1932, moyal1949}, which describes the single-particle density operator $\varrho(t)$ in terms of the Wigner quasi-probability distribution $W_t (x,p)$, see Appendix\,\ref{app:wigner} for the details. The variables $x$ and $p$ are canonically conjugated and depend on the form of the Hamiltonian, for Hamilton\,\eqref{eq:mathcalHQR} one can choose $\theta$ and $L$. 

The evolution equation of $W_t (x,p)$ is given by
\begin{equation} \label{eq:Moyal}
    \partial_t W_t (q,p) =  \{ \{ \mathcal{H}_{W} (t), W_t (q,p) \} \} \equiv - i \mathcal{L}_{W,\mathcal{H}} (W_t)
\end{equation}
with $\mathcal{H}_W = H^a_{0, W} - \lambda \mu^a(t) H^a_{1, W}$ and
\begin{equation}
    \mu^a(t) = \int dq dp \ H^a_{1, W} (q,p) W_t (q,p)   \, . 
\end{equation}
Here $\{ \{ \cdot, \cdot \} \}$ denote the Moyal brackets \cite{moyal1949}, while $H_{0,W}$ and $H^a_{1,W}$ denote the Weyl transform of $H_{0,W}$ and $H^a_1$ respectively,see Appendix\,\ref{app:wigner}. By comparing Eq.\,\eqref{eq:Moyal} with Eq.\,\eqref{eq:vonneuman} for any fixed value $\bar{\boldsymbol{\mu}}$ of $\boldsymbol{\mu}$, we see that the spectrum of $\mathcal{L}_{W,\mathcal{H}}$ must coincide with the one of $\mathcal{L}_H (\cdot) = [\mathcal{H}, \cdot]$, namely $\lbrace \mathcal{E}-\mathcal{E'} \rbrace$, where both $\mathcal{E}$ and  $\mathcal{E'}$ belong to the spectrum of $\mathcal{H}$. 

Within our assumptions, $W_t (x,p)$ becomes a bona fide probability distribution in the classical limit, while the Moyal brackets reduces to Poisson brackets
\begin{equation}
    \lbrace \lbrace \cdot, \cdot \rbrace \rbrace \rightarrow \lbrace \cdot, \cdot \rbrace
\end{equation} 
so that Eq.\,\eqref{eq:Moyal} becomes 
\begin{equation} \label{eq:vlasov}
    \partial_t W_t (q,p) =  \{ \mathcal{H}_{W} (t), W_t (q,p) \}  \equiv - i \mathcal{M}_{\mathcal{H}} (W_t) \, . 
\end{equation}
This can be straightforwardly interpreted as a Vlasov equation \cite{Vlasov1968, campa2014physics, giachetti2020coarse} for the mean-field evolution of a set of classical particles distributed in the phase space as $W_t (q,p) $. It is now clear that $\mathcal{M}_{\mathcal{H}}$ plays the role of the classical analogous of $\mathcal{L}_{H}$. Remarkably, it is possible to derive a general expression for the spectrum of $\mathcal{M}_{\mathcal{H}}$, and to show that it met the necessary condition of Sec.\,\eqref{sec:Condition} for VR.

Indeed, let us notice that Eq.\,\eqref{eq:vlasov} is covariant under canonical transformation of the canonical phase-space pair $x,p$. For any given (time-independent) $\mu$ it is thus possible to change variables to the action-angle pair $\phi,J$ so that $\mathcal{H}_W(x,p)$ becomes a function $\mathcal{H}_W (J)$ of $J$ alone. Eq.\,\eqref{eq:vlasov} becomes thus
\begin{equation}
    \partial_t W_t (\phi,J) = - \omega(J) \partial_\phi  W_t (\phi,J) \ ,  
\end{equation}
with $\omega(J) \equiv \mathcal{H}_W^{\prime} (J)$, and finally
\begin{equation}
    \mathcal{M}_{\mathcal{H}} = i \omega(J) \partial_\phi \ . 
\end{equation}
In turn, $\mathcal{M}_{\mathcal{H}}$ can be diagonalized in terms of $F_{n,J}$ are given by
\begin{equation}
    F_{n,J} (J^\prime,n) = e^{-in \phi} \delta(J-J^\prime) \ .
\end{equation}
with $n \in \mathbb{Z}$, as 
\begin{equation}
     \mathcal{M}_{\mathcal{H}} (F_{J,n}) =  n \omega(J) F_{J,n} \ .
\end{equation}
As $\omega(J)$ is, in general, a continuous function of $J$, the spectrum 
\begin{equation} \label{eq:cspectrum}
    \lbrace \Omega \rbrace = \lbrace n \omega(J) | n \in \mathbb{Z} \rbrace
\end{equation}
of $\mathcal{M}_{\mathcal{H}}$ exhibits a band structure for any $n \neq 0$, while for $n=0$ the whole band collapses into $\Omega = 0$.  By comparing it with the quantum version \eqref{eq:qspectrum} of the spectrum of the operator $\mathcal{L}_{\mathcal{H}}$, we see that $\lbrace \Omega \rbrace$ has exactly the structure expected from a quantum system with a single discrete point. Let us notice that a possible exception to this general behavior is given by the case in which $\mathcal{H}_W$ is quadratic, since in this case $\omega(J)$ is constant: in this case, however, it is known that VR cannot take place\,\cite{giachetti2019violent,giachetti2020coarse}.

To explain the physics behind this reasoning, let us consider the case of the quantum rotor Hamiltonian, examined in Sec.\,\ref{sec:HMF}. As already pointed out, in the $\hbar \rightarrow 0$ limit, the model reproduces the HMF model\,\cite{antoni1995clustering}, and the single-particle Hamiltonian \eqref{eq:mathcalHQR} becomes the Hamiltonian of a simple pendulum with a time-depending coupling $\mu(t)$. For any constant value of $\mu(t)\equiv \bar{\mu}$, the classical dynamics is confined on the level curves of the Hamiltonian. On the other hand, since the frequencies of such trajectories is not constant, bur rather it depends continuously on its energy, the phases of the different oscillations will become uncorrelated. At large times, thus, only the average value (i.e. the zero-th mode) over the classical trajectory will contribute to the evolution of observables, leading to VR.  

We can thus conclude that, in the classical limit, fully-connected mean-field system behaves effectively as a quantum system with a single (infinitely degenerate) element in the discrete spectrum. As a consequence, the necessary condition for the VR of Sec.\,\ref{sec:Condition} is generally met, explaining the ubiquitousness of the phenomenon in the classical world. 

\section{Conclusions}
\label{sec:conclusions}
In this paper we have explored the possibility for a quantum model with attractive, fully-connected, interactions, to undergo violent-relaxation (VR). While VR is rather common in the classical realm, this phenomenon is generally hindered in the quantum regime due to the presence of discrete ``bound states" in the spectrum of the effective, mean-field, Hamiltonian. If two or more discrete elements are present in the Hamiltonian spectrum, resonances between these states produce persistent Rabi oscillations and prevent VR. 

We test our predictions on a newly introduced spin Hamiltonian\,\eqref{eq:Hw}, characterized, in the large-$s$ limit, by a tunable number of bound states. Moreover, we verified that VR is absent in the two-component quantum rotor model\,\cite{sachdev1999quantum} \eqref{eq:HQR} despite reducing to the paradigmatic Hamiltonian-mean-field model\,\cite{antoni1995clustering} in the classical limit. Finally, we show how the necessary conditions for VR arise in the classical limit even for quantum systems with multiple bound states.

While the analysis is carried out in the formalism of commuting lattice variables, the fact that the effective Hamiltonian\,\eqref{eq:mathcalHQR} reproduces the Gross-Pitajevskij formalism discussed in Refs.\,\cite{plestid2018violent,chavanis2011quantumII} suggests that our picture can be extended to the bosonic case. It is worth noting that Hamiltonian\,\eqref{eq:mathcalHQR} has purely attractive interactions and our findings differ from the ones depicted in Ref.\,\cite{plestid2018violent}, where bi-clustering phenomena have been discussed.

All together, our results indicate that, regardless of the mean-field nature of the model, quantum effects play a crucial role in the dynamics and can alter the dynamical behavior at a qualitative level. As cold atoms into cavity are known to give raise to all-to-all effective interactions, our prediction could be verified experimentally in atomic, molecular, optical setups. On the other hand, the possibility of observing VR in quantum model could provide a new way for engineering stationary states in quantum technological applications.

\section{Acknowledgements}
We acknowledge funding by the Swiss
National Science Foundation (SNSF) under project funding
ID: 200021 207537 and by the Deutsche Forschungsgemeinschaft (DFG, German Research Foundation) under Germany’s
Excellence Strategy EXC2181/1-390900948 (the Heidelberg
STRUCTURES Excellence Cluster) and by the European
Union under GA No. 101077500–QLR-Net. This research
was supported in part by grant NSF PHY-230935 to the Kavli
Institute for Theoretical Physics (KITP).

\appendix
    
\onecolumngrid

\section{Mean-field limit}
\label{app:meanfield}
\noindent
In this section, we are going to present a possible derivation of the mean-field approximation.  Given a generic state $\rho(t)$ its evolution equation can be written as
\begin{equation} \label{eq:drhot}
\begin{split}
    i \frac{d}{dt} \rho = \sum_j [H_{0} (\boldsymbol{\tau}_j) , \rho] - \frac{\lambda}{2N} \sum_{j,j^{\prime}}  \big( H^a_1(\boldsymbol{\tau}_{j'}) [H^a_1(\boldsymbol{\tau}_j),\rho] + [H^a_1(\boldsymbol{\tau}_j),\rho] H^a_1(\boldsymbol{\tau}_{j'}) \big)
    \end{split} .
\end{equation}
Given two sites $x$,$x'$ with $x \neq x'$, let us now consider the quantities 
\begin{equation}
    \rho_x \equiv \operatorname{tr}_x \rho \, ,  \hspace{1cm}
    \rho^c_{xx'} \equiv \operatorname{tr}_{x,x'} \rho - \rho_x \otimes \rho_{x'} \, , 
\end{equation}
where $\operatorname{tr}_A$ denotes the partial trace on the Hilbert space of all the sites $ \notin A$. We have that $\rho^c_{xx'} = 0$ for any $x \neq x'$ if and only if the state $\rho$ is completely factorized. On the other hand, we are going to show that Eq.\,\eqref{eq:drhot}, in the thermodynamic limit, implies that $\dot{\rho}_{x,x'} = 0$ whenever the state is factorized. As a consequence, since at $t=0$ this is the case, the state is going to stay factorized at any time. 
\\\\
In order to show this result, let us notice that on a factorized state
\begin{equation}
    \operatorname{tr}_{x} \left( H^a_1(\boldsymbol{\tau}_{j}) \rho \right ) = \operatorname{tr} \left( H^a_1(\boldsymbol{\tau}_{j}) \rho \right ) \rho_x
\end{equation}
for any $x \neq j$ and 
\begin{equation}
\begin{split}
    \operatorname{tr}_x \big(H^a_1(\boldsymbol{\tau}_{j'}) [H^a_1(\boldsymbol{\tau}_{j}),  \rho] \big) &= \delta_{j,x} \operatorname{tr} ( H^a_1(\boldsymbol{\tau}_{j'}) \rho ) [ H^a_1(\boldsymbol{\tau}_{x}), \rho_x] \, , \\
    \operatorname{tr}_{x,x'} \big(H^a_1(\boldsymbol{\tau}_{j'}) [H^a_1(\boldsymbol{\tau}_{j}),  \rho] \big) &= \big(\delta_{j,x} \operatorname{tr}_{x'} \left( H^a_1(\boldsymbol{\tau}_{j'}) \rho \right) + \delta_{j,x'} \operatorname{tr}_{x} \left( H^a_1(\boldsymbol{\tau}_{j'}) \rho \right) \big) \otimes [H^a_1(\boldsymbol{\tau}_{j}),  \rho_{j}] \, , 
\end{split}
\end{equation}
for any $j \neq j'$. As a consequence
\begin{equation}
\begin{split}
    i \frac{d}{dt} \rho_x &=  [H_{0} (\boldsymbol{\tau}_x)  - \lambda \mu^a (t) H^a_1(\boldsymbol{\tau}_x),\rho_x] + O(N^{-1}) \, , \\
    i \frac{d}{dt} \operatorname{tr}_{x,x'} \rho &=  \rho_x \otimes [H_{0} (\boldsymbol{\tau}_{x'}) -  \lambda \mu^a (t) H^a_1(\boldsymbol{\tau}_{x'}), \rho_{x'}] + (x \longleftrightarrow x') + O(N^{-1})  \, , 
\end{split} 
\end{equation}
with 
\begin{equation}
    \mu^a(t) = \frac{1}{N} \sum_{j}  \operatorname{tr} (H^a_1(\boldsymbol{\tau}_{j}) \rho ) \, ,  
\end{equation}
and the $O(N^{-1})$ terms comes from the terms with $j' = j$, $j = x,x'$ and $j'=x,x'$ in the second sum of the right-hand-side of Eq.\,\eqref{eq:drhot}. It follows that, in the $N \rightarrow \infty$ limit
\begin{equation}
    \frac{d}{dt} \rho^c_{x,x'} (t) = 0 \, , 
\end{equation}
while for finite $N$ those correlation will build up on a time polynomial in $N$. 
\\\\
We can thus conclude that $\mu(t)$, being the expectation value of the a sum of single-site operators, can be written as $\mu^a (t) = \operatorname{tr} H^a_1 \varrho(t) \, $, where the effective single-site density operator $\varrho = 1/N \sum_j \rho_j$ evolves with 
\begin{equation} \label{eq:varrhoevolution}
   i \frac{d}{dt} \varrho = [H_0 (\boldsymbol{\tau}) - \lambda \mu^a (t) H^a_1 (\boldsymbol{\tau}), \varrho] \, ,  
\end{equation}
which is exactly the mean-field picture exposed in the text. From this analysis it also follows that the energy density $\epsilon = \tr(\varrho(t) H_0) - \lambda \boldsymbol{\mu}^2(t)/2$ is conserved. Indeed
\begin{equation}
        \dot{\epsilon} = \tr(\dot{\varrho} H_0) - \lambda \mu^a  \tr(\dot{\varrho} H^a_1) = \tr(\dot{\varrho} \mathcal{H}) =  \tr([\mathcal{H},\varrho] \mathcal{H}) = 0 
\end{equation}

\section{Spectral properties of the \emph{w}-model}
\label{app:wspectrum}
\noindent
We will now examine the structure of the spectrum of the singe Hamiltonian \eqref{eq:mathcalHw}
\begin{figure}
    \centering
    \includegraphics[width= 0.6
    \textwidth]{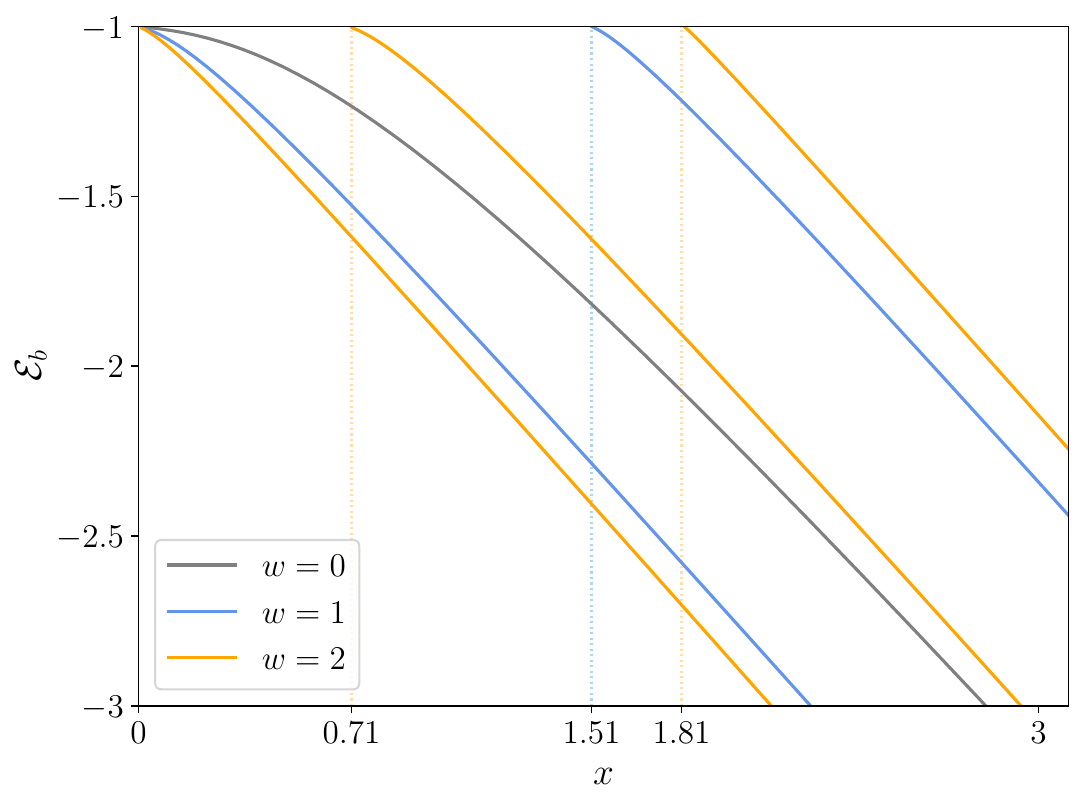}
    \caption{Even bound states of the Hamiltonian $\mathcal{H}_w$ as function of $x$, for $w=0,1,2$, obtained by exact diagonalizing the matrix \eqref{eq:mathcaHwmm'} for $s=500$ and selecting only the eigenvalues $\leq -1$ corresponding to an even eigenstate. While, regardless of $w$, a bound state is present for any value of $x$, for $w>0$ new even bound states appear in correspondence of critical values of $x$ ($x_c \sim 1.51$ for $w=1$, $x_c \sim 0.71, 1.81$ for $w=2$). The maximum number of even bound states is $w+1$.}
    \label{fig:SpectrumW}
\end{figure}
\begin{equation} 
     \mathcal{H}_w (t) = - \frac{h}{s} s_x - \lambda \bar{\mu} \theta_H (w^2 - s_z^2) ,
\end{equation}
for fixed $\bar{\mu}$. In the following we will choose $h=1$. 

Let us choose the basis of eigenstates of $s_z$, $\lbrace \ket{m}  \rbrace$ with $m = -s, \dots, s$. Then 
\begin{equation} \label{eq:mathcaHwmm'}
    \bra{m'} \mathcal{H}_w \ket{m} = - \frac{1}{2s} \sqrt{s(s+1)-m'm} \left( \delta_{m,m'+1} + \delta_{m,m'-1} \right) - x \delta_{m'm} \theta_H (w^2-m^2) 
\end{equation}
where we introduced $x = \lambda \bar{\mu} > 0$. The $m \rightarrow - m$ invariance of $\mathcal{H}_w$ is a consequence of the $\mathbb{Z}_2$ symmetry of the model. For $x=0$ the spectrum of the Hamiltonian
\begin{equation}
     \lbrace \mathcal{E}  \rbrace = \lbrace p/s \ | p = -s, \dots, s \rbrace
\end{equation}
approaches a continuum uniform density in the interval $[-1,1]$. For any $x > 0$ discrete points, with $\mathcal{E} < -1$, appear in the spectrum, which can be interpreted as bound states, in which the eigenstate is localized around $m=0$ because of the presence of the well potential. We will thus refer to them as bound states $\lbrace \mathcal{E}_b \rbrace$. To estimate the number of bound states let us notice that, for $|m| \leq w$ and $s \rightarrow \infty$ (with $w$ fixed) we have
\begin{equation}
    \bra{m'} \mathcal{H}_w \ket{m} = - \frac{1}{2} \left( \delta_{m,m'+1} + \delta_{m,m'-1} \right) - x \delta_{m'm} \theta(w^2 - m^2) \ ; 
\end{equation}
if we now consider the $x \rightarrow \infty$ limit the eigenvectors will vanish for $|m| > w$, so that we can write them as
\begin{equation}
    \braket{m|\psi_p} = e^{\pi i m p/(2w+2)} - (-1)^p  e^{-\pi i m p/(2w+2)}
\end{equation}
with $p= 1, \dots, 2w+1$ (the corresponding eigenvalues are given by $\mathcal{E}_b = \lbrace \cos(\pi p/(2w+2)) - x \rbrace$). It follows that for finite $x$ the maximum number of bound states is $2w+1$. Taking into account the $\mathbb{Z}_2$ symmetry, we have a maximum of $w+1$ even bound states and $w$ odd states. This is confirmed by the numerical analysis shown in Fig.\,\ref{fig:SpectrumW} in which the even bound eigenstate energies are plotted as a function of $x$ for $w=0,1,2$. for $w>0$ new bounds states appear in correspondence of critical values $x_c$ of $x$ (in particular, for $w=1$, $x_c \sim 1.51$).

\section{Wigner function formalism}
\label{app:wigner}
\noindent
In this paragraph we will briefly revisit the Wigner function formalism, which is a standard tools for considering the classical limit of quantum systems. Given a set of canonically conjugate quantum variables $x$,$p$, $[x,p]=i$; we consider the associated phase space in which $x$ and $p$ play the role of coordinates. The state $\varrho(t)$ of the system can be associated to the so-called Wigner function 
\begin{equation}
    W_t (x,p) = \frac{1}{\pi} \int dy \ \bra{x-y} \varrho (t) \ket{x+y} e^{2ipy} \ , 
\end{equation}
which is normalized 
\begin{equation}
    \int dx dp \ W_t(x,p) = 1
\end{equation}
but in general non positive defined, so that $W_t$ is a quasi-probability distribution. Similarly, given a generic observable $A$, we can define its Weyl transform 
\begin{equation}
    A_W (x,p) =  2 \int dy \ \bra{x-y} A \ket{x+y} e^{2ipy} \ . 
\end{equation}
It can be shown that 
\begin{equation}
    \me{A} = \tr (\varrho A) = \int dx dp \ A_W(x,p) W_t(x,p) 
\end{equation}
The time-evolution of the Wigner function is given by 
\begin{equation}
     \partial_t W_t(x,p) = \{ \{ H_W, W_t \} \}
\end{equation}
where $\lbrace \lbrace \cdot, \cdot \rbrace \rbrace$ denotes the Moyal brackets
\begin{equation}
    \lbrace \lbrace A, B \rbrace \rbrace = 2 A(x,p) \sin \left( \frac{1}{2}  (\overleftarrow{\partial_x} \overrightarrow{\partial_p} - \overleftarrow{\partial_p} \overrightarrow{\partial_x} )\right) B(x,p) \ . 
\end{equation}
In particular, the evolution of the state $\varrho(t)$ in the mean-field limit will be given by 
\begin{equation} \label{eq:moyalevolution}
     \partial_t W_t = \lbrace \lbrace \mathcal{H}_W, W_t \rbrace \rbrace = \lbrace \lbrace \mathcal{H}_{0,W} - \lambda \mu(t) \mathcal{H}_{1,W} , W_t \rbrace \rbrace 
\end{equation}
where 
\begin{equation}
    \mu(t) = \int dx dp \ W_t (x,p) \mathcal{H}_{1,W} (x,p) \ .
\end{equation}

\twocolumngrid

\end{document}